\documentclass[12pt]{iopart}

\usepackage{iopams}  
\usepackage[dvips,dvipdfm]{graphicx}

\usepackage{amsfonts}

\newcommand{\bn}{\mbox{\boldmath{$n$}}}

\newcommand{\bbf}{\mbox{\boldmath{$x$}}}

\newcommand{\bbs}{\mbox{\boldmath{$y$}}}

\newcommand{\bx}{\mbox{\boldmath{$x$}}}

\newcommand{\ba}{\mbox{\boldmath{$a$}}}

\newcommand{\by}{\mbox{\boldmath{$y$}}}

\newcommand{\mR}{\mathbb{R}}
\newcommand{\mN}{\mathbb{N}}

\begin{document}

\title[Typical performance of compressed sensing ]{A typical reconstruction limit of compressed sensing based on 
$L_p$-norm minimization }
\author{Y. Kabashima$^1$, T. Wadayama$^2$ and T. Tanaka$^3$}

\address{$^1$
Department of Computational Intelligence and Systems Science, 
Tokyo Institute of Technology - Yokohama 226-8502, Japan \\
$^2$
Department of Computer Science, 
Nagoya Institute of Technology - Nagoya 466-8555, Japan \\
$^3$
Department of Systems Science, 
Kyoto University - Kyoto 606-8501, Japan
}

\begin{abstract}
We consider the problem of reconstructing an $N$-dimensional 
continuous vector $\bbf$ from $P$ constraints 
which are generated by its linear transformation under the 
assumption that the number of non-zero elements of $\bbf$ 
is typically limited to $\rho N$ ($0\le \rho \le 1$). 
Problems of this type can be 
solved by minimizing a cost function 
with respect to the $L_p$-norm $||\bbf||_p=
\lim_{\epsilon \to +0}\sum_{i=1}^N
|x_i|^{p+\epsilon}$, subject to the 
constraints under an appropriate condition. 
For several $p$, we assess a typical case limit $\alpha_c(\rho)$, 
which represents a critical relation between $\alpha=P/N$ and $\rho$
for successfully reconstructing the original vector by minimization 
for typical situations in the limit $N,P \to \infty$ 
with keeping $\alpha$ finite, utilizing the replica method. 
For $p=1$, $\alpha_c(\rho)$ is considerably smaller than its worst case 
counterpart, 
which has been rigorously derived by existing literature 
of information theory. 
\end{abstract}

\maketitle

\section{Introduction}
{\em Compressed (or compressive) sensing }is a technique 
for reconstructing a high dimensional signal from lower dimensional data, the 
components of which represent partial information about the signal, utilizing 
prior knowledge on the sparsity of the signal. 
The research history of this technique is rather long 
\cite{Claerbout,Santosa,Donoho1}; but 
the horizon of the research field is now expanding rapidly 
after recent publication of a series of influential papers
\cite{Donoho2,CandesTao3,CandesTao1,CandesTao2}. 

In a recent paper, the following issue has been considered \cite{CandesTao3}. 
Let us suppose 
a situation where an $N$-dimensional continuous 
signal $\bbf \in \mR^N$ is compressed to a vector of
dimension $P(<N)$, $\bbs \in \mR^P$, utilizing a $P \times N$ 
signal-independent
compression matrix $F \in \mR^{P \times N}$ as 
\begin{eqnarray}
\bbs=F\bbf. 
\label{check}
\end{eqnarray}
We also assume that 
$F$ is known and that $\bbf$ is sparse in the sense that
the number of non-zero elements of $\bbf$
is limited to $\rho N$, where $0\le \rho  \le 1$. 
Then, under what conditions can the original 
signal $\bbf$ be correctly reconstructed 
from the compressed expression $\bbs$?

It is obvious that eq.~(\ref{check}) in itself cannot
determine a unique solution of $\bbf$ because the 
dimension of $\bbs$, $P$, is smaller than that of $\bbf$, $N$. 
However, the assumption on the sparsity of $\bbf$ may allow 
correct reconstruction. 
In the research on compressed sensing, 
minimization of a cost function with respect to 
the $L_p$-norm\footnote{Eq.~(\ref{Lp-norm}) does not define 
a norm in the mathematical sense 
because it violates the triangle inequality.}
\begin{eqnarray}
||\bbf||_p&=&
\lim_{\epsilon \to +0}
\sum_{i=1}^N|x_i|^{p+\epsilon}\cr
&=&
\left \{
\begin{array}{ll}
\sum_{i=1}^N |x_i|^p, & p > 0, \cr
\mbox{the number of non-zero elements of $\bbf$}, &p=0, 
\end{array}
\right .
\label{Lp-norm}
\end{eqnarray}
subject to the constraints of 
eq.~(\ref{check}) 
has been actively studied 
toward designing efficient reconstruction schemes
exploiting such sparsity
\cite{Danzig,HauptNowak,Lasso,ChenDonohoSaunders}.

Results of \cite{CandesTao3,CandesTao2,Candes2008}
indicate that the following proposition holds. 
Let us suppose that $\bbf$ is an {\em arbitrary} continuous real vector 
the number of non-zero elements of which is bounded above by $S$. 
When each entry of $F$ is an independently and identically distributed 
(i.i.d.) Gaussian random number, the probability 
of failure in reconstructing 
$\bbf$ based on the $L_1$-norm minimization 
becomes arbitrarily small 
as $N$ tends to infinity
if the inequalities
\begin{eqnarray}
\frac{2S}N\ln \left (\frac{N}{2S} \right )+\frac{2S}{N}+\frac{1}{N}
\ln (2S)-\frac{P}{2N}\left 
(2^{1/4}-1-\sqrt{\frac{2S}{P}}\right )^2 < 0, 
\label{sufficient1}
\end{eqnarray}
and
\begin{eqnarray}
2^{1/4}-1-\sqrt{\frac{2S}{P}} > 0, 
\label{sufficient2}
\end{eqnarray}
hold simultaneously. 
These inequalities constitute a sufficient condition for 
arbitrarily reducing the probability of failure for 
the $L_1$-based reconstruction 
of the arbitrary vector $\bbf$.
However, earlier studies 
on several other problems
in information theory indicate that 
critical conditions of such {\em worst cases} \cite{Gilbert,VC} are, 
in general, considerably different from those of {\em typical cases}
\cite{Shannon,Cover}, and are not necessarily 
relevant in practical situations. 

This Letter is written from such a perspective. 
More precisely, we will herein assess a critical 
condition for successfully reconstructing $\bbf$
in typical cases in the limit 
$N,P \to \infty$, but keeping $\alpha=P/N$ finite,
utilizing methods of statistical mechanics. 
Results of numerical experiments reported in \cite{CandesTao3} 
indicate that a critical condition of the reconstruction success 
for typical cases is far from that of eqs.~(\ref{sufficient1})
and (\ref{sufficient2}). 
Our result is in excellent agreement 
with this indication.

\section{Problem setting}
For simplicity, we will hereinafter assume the following. 
Each component of the original signal $\bbf^0 \in \mR^N$, 
$x_i^0$ ($i=1,2,\ldots,N$), is independently and identically generated from 
the distribution 
$P(x)=(1-\rho)\delta(x)+\rho \exp (-x^2/2 )/\sqrt{2 \pi}$, 
where 
$\rho$ ($0\le \rho \le 1$) is referred to as {\em signal density}
and $\delta(u)$ is Dirac's delta function. 
The compressed expression $\bbs$ is 
provided as $\bbs=F\bbf^0$.
We assume that $\bbs$ and $F$ are available
but $\bbf^0$ is hidden. 
Each entry of $F$, $F_{\mu i}$ ($\mu=1,2,\ldots,P$; $i=1,2,\ldots,N$), 
is an i.i.d.\ Gaussian random variable of mean zero and 
variance 
$N^{-1}$. 

For generality, we formally consider a general 
reconstruction scheme 
\begin{eqnarray}
{\rm minimize} \ ||\bbf||_p \
{\rm subject \ to }\
F\bbf = \bbs, 
\label{Lp}
\end{eqnarray}
utilizing a cost function with respect to the $L_p$-norm. 
We will refer to eq.~(\ref{Lp}) as {\em $L_p$-reconstruction}. 
In the following, we will generally examine the typical
reconstruction performance for the cases of $p=0,1$ and $2$ 
in the limit $N,P \to \infty$, 
but keeping 
{\em compression rate }
$\alpha=P/N$ finite. 

In a recent work, utility of 
the $L_p$-norm cost function in estimating $\bbf^0$ from $\bbs+\bn$
is examined, where $\bn$ is 
a zero mean Gaussian noise vector \cite{Rangan}. 
In such problem setting, however, 
correct reconstruction of $\bbf^0$, which 
we will focus on hereinafter, is not possible 
as long as the variance per element of $\bn$ is finite. 

\section{Analysis}
To directly assess the typical performance of the $L_p$-reconstruction, 
we have to solve eq.~(\ref{Lp}) and examine whether
the solution that is obtained is identical to $\bbf^0$ or not
for {\em each} sample of randomly generated $F$ and $\bbs(=F\bbf^0)$. 
Carrying this out analytically is, unfortunately, difficult in practice. 
To avoid this difficulty, we convert the constrained 
minimization problem of eq.~(\ref{Lp}) to a posterior distribution 
of the inverse temperature $\beta$, thus: 
\begin{eqnarray}
P_{\beta}(\bbf|\bbs)=\frac{e^{-\beta ||\bbf||_p} 
\delta\left (F\bbf-\bbs \right )}{Z(\beta;\bbs)}, 
\label{boltzmann}
\end{eqnarray}
where $Z(\beta;\bbs)=\int d\bbf e^{-\beta ||\bbf||_p} 
\delta\left (F\bbf-\bbs \right )$ plays the role of a partition function. 
In the limit $\beta \to \infty$, eq.~(\ref{boltzmann}) generally converges 
to a uniform distribution over the solutions of eq.~(\ref{Lp}). 
Therefore, one can evaluate the performance of 
the $L_p$-reconstruction scheme by examining the macroscopic 
behavior of eq.~(\ref{boltzmann}) as $\beta \to \infty$, 
for which one can utilize methods of statistical mechanics. 

A distinctive feature of the current problem is that 
eq.~(\ref{boltzmann}) depends on 
the predetermined (quenched)
random variables $F$ and 
$\bbf^0$ 
(through $\bbs=F\bbf^0$), 
which naturally leads us to applying the replica method \cite{Dotzenko}. 
Under the replica symmetric (RS) ansatz, this yields 
an expression of the typical free energy density 
as $\beta \to \infty$ as
\begin{eqnarray}
C_p&=&-\lim_{\beta \to \infty}\lim_{N \to \infty}
\frac{1}{\beta N} \left [
\ln Z(\beta;\bbs) \right ]
=-\lim_{\beta \to \infty}
\lim_{n \to 0}
\frac{\partial }{\partial n}
\lim_{N \to \infty}
\frac{1}{\beta N} \ln \left [Z^n(\beta;\bbs) \right ] \cr
&=&\mathop{\rm extr}_{\Theta}
\left \{
\frac{\alpha(Q-2m+\rho)}{2 \chi}+\widehat{m}m-\frac{\widehat{Q}Q}{2}
+\frac{\widehat{\chi}\chi}{2} \right .
\cr
&\phantom{=}&
\left .
+(1-\rho)\int Dz \phi_p\left (\sqrt{\widehat{\chi}} z;\widehat{Q} \right)
+\rho\int Dz \phi_p
\left (\sqrt{\widehat{\chi}+\widehat{m}^2}z;\widehat{Q} \right ) 
\right \}, 
\label{freeenergy}
\end{eqnarray}
where 
$\left [\cdots \right ]$ represents 
the operation of averaging with respect to $F$ and 
$\bbf^0$, 
and $\mathop{\rm extr}_{X}\{ {\cal G}(X) \}$ denotes
extremization of a function ${\cal G}(X)$ with respect to $X$, 
$\Theta=\{Q,\chi,m,\widehat{Q},\widehat{\chi},\widehat{m}\}$, 
$Dz=dz \exp (-z^2/2 )/\sqrt{2 \pi}$ is a Gaussian measure and 
\begin{eqnarray}
\phi_p(h;\widehat{Q})
=\lim_{\epsilon \to +0}
\left \{\mathop{\rm min}_{x}\left \{
\frac{\widehat{Q}}{2}x^2-h x+|x|^{p+\epsilon}
\right \} \right \}. 
\label{phi_p}
\end{eqnarray}
The term $\mathop{\rm min}_{X}\left \{{\cal G}(X) \right \}$ denotes 
minimization of ${\cal G}(X)$ with respect to $X$. 
A sketch of the derivation is shown in \ref{derivation}. 

\begin{figure}[t]
\centerline{\includegraphics[width=12cm]{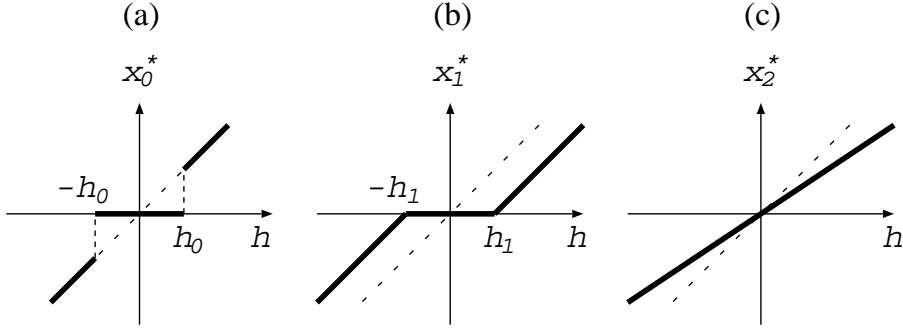}}
\caption{Profiles of $x_p^*(h;\widehat{Q})$ for
$p=0,1$ and $2$. 
(a): $x_0^*(h;\widehat{Q})=h/\widehat{Q}$ for $|h| > h_0$ and 
$0$, otherwise, where $h_0=\sqrt{2 \widehat{Q}}$. 
(b): $x_1^*(h;\widehat{Q})=(h-h/|h|)/\widehat{Q}$ for $|h| > h_1$ and 
$0$, otherwise, where $h_1=1$. 
(c): $x_2^*(h;\widehat{Q})=h/(\widehat{Q}+2)$. 
}
\label{fig1}
\end{figure}

Three issues are noteworthy here. 
The first issue concerns the physical meanings of the
variables introduced in eq.~(\ref{freeenergy}). 
For example, at the extremum, 
values of $Q$ and $m$ 
in eq.~(\ref{freeenergy}) correspond to 
$N^{-1} \left [ \left \langle |\bbf \right |^2 \rangle\right ]$
and $N^{-1} \left [ \bbf^0 \cdot \left \langle \bbf \right \rangle \right ]$, 
respectively, where $\left \langle \cdots \right \rangle$
denotes averaging with respect to eq.~(\ref{boltzmann}) 
as $\beta \to \infty$ and 
$|\ba|$ denotes the ordinary Euclidean norm
$\sqrt{\sum_{i} |a_i|^2}$
for a vector $\ba=(a_i)$. 
This indicates that 
the typical value of the mean square error per component 
${\rm MSE}=N^{-1} \left [\left \langle 
|\bbf-\bbf^0|^2 \right \rangle \right ]$
can be assessed as
\begin{eqnarray}
{\rm MSE}=Q-2m+\rho, 
\label{MSE}
\end{eqnarray}
utilizing the extremum solution of eq.~(\ref{freeenergy}). 
When the correct signal $\bbf^0$ dominates eq.~(\ref{boltzmann})
as $\beta \to \infty$, $Q=m=N^{-1}\left [ |\bbf^0|^2 \right ]=\rho$
holds. Therefore, one can argue the typical possibility of 
correct reconstruction by examining whether this success solution 
dominates the extremization problem of eq.~(\ref{freeenergy}) or not. 
In addition, the extremized value of eq.~(\ref{freeenergy}),
$C_p$, itself also possesses 
the physical meaning of a typical value
of the minimized $L_p$-norm (per element) as
$C_p=N^{-1} \sum_{i=1}^N
\lim_{\epsilon \to +0}\left [ \left \langle 
 |x_i|^{p+\epsilon} \right \rangle \right ]$. 
The second concerns the practical implication of eq.~(\ref{phi_p}). 
Comparison with analysis of the cavity method
\cite{MezardParisiVirasoro}, which is an alternative to the 
replica method, indicates that eq.~(\ref{phi_p}) stands for 
the minimization problem that a single site is 
effectively required to solve when the site is newly added to a cavity 
system which is defined by removing the site out of the original system. 
In that situation, the two terms $\widehat{Q}f^2/2-h f$ constitute 
an effective cost which arises from the constraints of eq.~(\ref{check}) 
by taking an average with respect to 
eq.~(\ref{boltzmann}) of the cavity system. 
Fig.~\ref{fig1} shows how the optimal solution 
of the right-hand side of eq.~(\ref{phi_p}) 
given $h$ and $\widehat{Q}$, 
denoted by $x_p^*(h;\widehat{Q})=-
\partial \phi_p(h;\widehat{Q}) /\partial h$, 
behaves for $p=0,1$ and $2$. 
The function $x_p^*(h;\widehat{Q})$ is of utility 
for constructing an approximation algorithm 
to solve eq.~(\ref{Lp}) given a single instance of 
$F$ and $\bbs$. 
The final issue concerns the validity of 
the RS solution of eq.~(\ref{freeenergy}). 
In applying the replica method to 
the current system, the generalized moment of the partition function
$\left [Z^n(\beta;\bbs) \right ]$ $(n \in \mR)$  is assessed
by analytically continuing the expression of the 
saddle point evaluation of $\left [Z^n(\beta;\bbs) \right ]=
\int \prod_{a=1}^n d\bbf^a 
e^{-\beta\sum_{a=1}^n ||\bbf^a||_p }
\left [ \prod_{a=1}^n \delta(F\bbf^a-\bbs) \right ]$
for $n \in \mN$ to $n \in \mR$. 
For such an assessment, it is generally required to introduce 
an assumption about how the dominant saddle point 
behaves under permutation with respect to the replica indices
$a=1,2,\ldots,n$. For deriving eq.~(\ref{freeenergy}), 
we have adopted the RS ansatz, in which the dominant saddle 
point is assumed to be invariant under any permutation of 
the replica indices. However, 
local stability of the RS saddle point 
fails with respect to perturbations that 
break the replica symmetry 
if 
\begin{eqnarray}
&& \frac{\alpha}{\chi^2}
\left ((1-\rho)
\int Dz \left (\frac{\partial x_p^*(\sqrt{\widehat{\chi}}z;\widehat{Q})}{
\partial (\sqrt{\widehat{\chi}}z)}
\right )^2 \right . \cr
&&\phantom{aaaaaaaaaaaaaaaaaa}
\left . 
+ \rho 
\int Dz \left (\frac{\partial x_p^*(\sqrt{\widehat{\chi}+\widehat{m}^2}z;\widehat{Q})}{
\partial (\sqrt{\widehat{\chi}+\widehat{m}^2}z)}
\right )^2 \right ) > 1, 
\label{AT}
\end{eqnarray}
holds, which represents the de Almeida-Thouless (AT) 
instability condition for the present problem \cite{AT}. 
When eq.~(\ref{AT}) holds for the extremum solution 
of eq.~(\ref{freeenergy}), the RS treatment is not valid 
and one has to explore more general solutions taking the effect of 
replica symmetry breaking (RSB) into account to accurately assess 
the performance of the $L_p$-reconstruction. 

\section{Results}
For $p=0,1$ and $2$, we numerically solved the RS extremization problem 
of eq.~(\ref{freeenergy}) for various pairs of $\alpha$ and $\rho$.  
In all cases, only a single stable solution was found. 
Given $\rho$ and $p$, the solution found 
for sufficiently large $\alpha$ was always characterized 
by $Q=m=\rho$ indicating successful reconstruction.  
However, as $\alpha$ was lowered, the success solution 
lost its local stability (against the RS disturbance) 
and a transition to a failure solution of $Q\ne m \ne \rho$ occurred. 

For the success solution, conjugate variables
$\widehat{Q}$ and $\widehat{m}$ were always infinitely large
whereas the remaining variables $\chi$ and $\widehat{\chi}$
did not 
necessarily
diverge. 
Investigating local stability of the success solution 
yielded a limit $\alpha_c(\rho)$, 
which represented the possibility of $L_p$-reconstruction in typical cases. 
For each of $p=0,1 $ and $2$, this is summarized as follows. 

\subsection{$p=0$}
The success solution, for which 
$\chi=0$ and 
$\widehat{\chi}\to \infty$, 
is stable if and only if
$\alpha > \rho$, which indicates $\alpha_c(\rho)=\rho$. 
The condition $\alpha > \rho$ is necessary to ensure that eq.~(\ref{check})
has a unique solution, even in the situation 
that all sites of non-zero elements of $\bbf$ are known. 
This means that the limit for $p=0$ achieves the best possible 
performance. 
However, due to discontinuity in the profile of $x_0^*(h;\widehat{Q})$
(Fig.~\ref{fig1} (a)), eq.~(\ref{AT}) always holds for the success 
solution, indicating that the current RS analysis is not valid. 
Therefore, further exploration based on 
various RSB ans\"{a}tze 
is necessary for accurately assessing 
the reconstruction performance, which is, however,
beyond the scope of the present Letter. 

\subsection{$p=1$} $\widehat{\chi}$ of the success solution is
determined by 
\begin{eqnarray}
\widehat{\chi}=\alpha^{-1}
\left [
2(1-\rho)\left (
(\widehat{\chi}+1)H(\widehat{\chi}^{-1/2})
-\widehat{\chi}^{-1/2}\frac{e^{-1/(2 \widehat{\chi})}}{\sqrt{2\pi}}
\right )+\rho(\widehat{\chi}+1) 
\right ], 
\label{determine_chi_hat}
\end{eqnarray}
where $H(x)=\int_x^{\infty}Dt$. 
Utilizing the solution of this equation, the stability condition of 
the success solution is expressed as
\begin{eqnarray}
\alpha > 2(1-\rho)H(\widehat{\chi}^{-1/2})+\rho. 
\label{L1capacity}
\end{eqnarray}
This indicates that 
the limit for $p=1$ can be 
expressed as $\alpha_c(\rho)=2(1-\rho)H(\widehat{\chi}^{-1/2})+\rho$. 
$\alpha_c(\rho)$ also 
corresponds to the criticality of eq.~(\ref{AT}) 
and the RS success solution is 
locally stable against perturbations that break the replica symmetry 
as long as eq.~(\ref{L1capacity}) holds. 
Therefore, our RS analysis is valid. 

\subsection{$p=2$}
The success solution is stable if and only if $\alpha \ge 1$, 
implying $\alpha_c(\rho)=1$. For $\alpha > \alpha_c(\rho)=1$, 
eq.~(\ref{AT}) does not hold and the RS analysis is valid. 
Since $\alpha \ge 1$ makes the constraints of eq.~(\ref{check})
sufficient to reconstruct $\bx^0$ perfectly, 
this result means that the $L_2$-norm minimization is not capable 
of reconstructing any compressed expressions. 

\begin{figure}[t]
\centerline{\includegraphics[width=15cm]{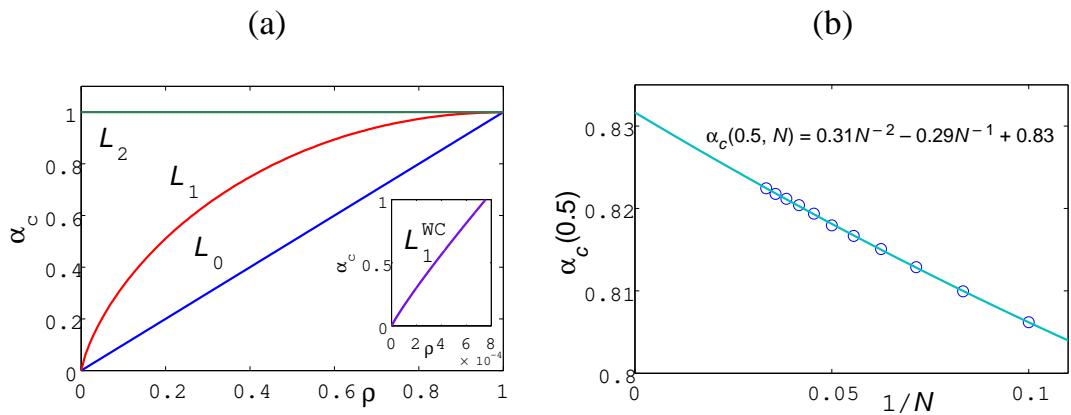}}
\caption{(a): Comparison of typical reconstruction limits of
the $L_p$-reconstruction for $p=0,1$ and $2$. 
Each curve denoted by ``$L_p$'' 
represents the RS estimate of the typical critical 
compression rate $\alpha_c(\rho)$
for 
the signal density $\rho$ of the original signal
for the $L_p$-reconstruction scheme. 
Correct reconstruction is typically possible for $\alpha > \alpha_c(\rho)$,
but the RS estimate for $p=0$ is not physically valid
due to the AT instability of eq.~(\ref{AT}). 
``$L_1^{\rm WC}$'' (inset) represents a worst case counterpart
for the $L_1$-reconstruction assessed from eqs.~(\ref{sufficient1})
and (\ref{sufficient2})
by setting $P=\alpha N$ and $S=\rho N$. 
(b): Experimental assessment of $\alpha_c(\rho=0.5)$
for the $L_1$-reconstruction. 
Experimental data of $\alpha_c(0.5,N)$ (see main text) 
for $N=10,12,\ldots,30$ was fitted by a quadratic function of $1/N$. 
This yields a value of extrapolation $\alpha_c(0.5) \simeq 0.83165$ 
for $N \to \infty$, which is close to the theoretical estimate
$\alpha_c(0.5)=0.83129\ldots.$ }
\label{fig2}
\end{figure}


Plots of the results obtained are shown in Fig.~\ref{fig2} (a). 
We also depict a curve of the worst case critical condition for 
the $L_1$-reconstruction (inset), 
which is assessed utilizing eqs.~(\ref{sufficient1}) 
and (\ref{sufficient2}) 
in the limit $N,P \to \infty$, keeping $\alpha=P/N$ 
and $\rho=S/N$ finite, for comparison. 
The $L_1$-reconstruction can be carried out in practice 
by interior point methods \cite{interior}, the necessary computational 
cost of which grows as $O(N^3)$ in the present large system limit. 
On the other hand, performing the $L_0$-reconstruction is, 
in general, NP hard, 
although its potential 
might be superior to that of the $L_1$-reconstruction. 
Fig.~\ref{fig2} (a), in conjunction with these, implies that 
the $L_1$-based scheme is a practically preferable method
which balances computational feasibility and relatively 
high reconstruction capability. 
This figure also indicates that discrepancy of the 
values of critical compression rate 
is huge between the worst and typical case
analyses.
This implies that there may be much room for improvement 
of the worst case assessment
although we must keep in mind that 
the criterion of reconstruction success in the present analysis, 
which permits reconstruction errors of asymptotically 
negligible size as $N \to \infty$, 
is different from that of the worst case analysis, 
in which no errors are allowed. 

To justify our assessment, we performed extensive 
numerical experiments of the $L_1$-reconstruction
for $\rho=0.5$, the results of which are summarized in Fig.~\ref{fig2} (b). 
In an experimental trial, an original signal $\bbf^0$ was 
randomly generated so as to have exactly $S=\rho N= N/2$ non-zero 
elements, to which i.i.d.\ Gaussian random numbers of zero mean 
and unit variance were assigned. For numerically assessing the criticality, 
the number of constraints $P$ was lowered from $P=N$ one-by-one 
until the solution of the $L_1$-reconstruction, 
$\widehat{\bbf}$, satisfied the condition of $||\widehat{\bbf}-\bbf^0||_1 >
10^{-4}$, and $P_c=P+1$ was recorded
when the condition was first satisfied. 
For searching for $\widehat{\bbf}$, we used \texttt{CVX}, a package 
for specifying and solving convex programs \cite{CVX,DCP}.
The trials were carried out $10^6$ times for a fixed system size $N$
and the experimental critical rate was defined as 
$\alpha_c(\rho=0.5,N)=\overline{P_c} /N$, where
$\overline{\cdots}$ denotes the arithmetic average over the trials.
Quadratic extrapolation from data for $N=10,12,\ldots,30$ yielded
an experimental estimate of the critical ratio $\alpha_c(0.5)=
\lim_{N \to \infty}\alpha_c(0.5,N)\simeq 0.83165$, 
which is in good accordance with the theoretical value
$\alpha_c(0.5)=0.83129\ldots$ (Fig.~\ref{fig2} (b)). 
In \cite{CandesTao3}, 
experiments for evaluating the critical density $\rho_c$
for $\alpha=0.5$ were performed for relatively large systems
of $N=512$ and $1024$. Judging from comparison by eye, 
plots of the results are also consistent with our theoretical estimate 
$\rho_c(\alpha=0.5)=0.19284\ldots$. 
These indicate that our assessment is at least capable of explaining 
the experimental results to a high accuracy 
although mathematical justification of the replica method, in general, 
has not yet been established \cite{Talagrand}.

\section{Summary and discussion}
In summary, we have assessed the typical performance of compressed sensing
based on minimization with respect to the $L_p$-norm 
for $p=0,1$ and $2$, utilizing 
the replica method under the replica symmetric (RS) ansatz. 
Analysis of the stability condition of a solution which 
represents successful reconstruction yields a critical 
relation between the compression rate and 
the signal density
that represents the frequency of non-zero elements in the original signal. 
We have shown that the RS solution of the 
$L_0$-reconstruction achieves the best possible performance, which 
is, unfortunately, not stable against perturbations that break the
replica symmetry. The $L_2$-reconstruction has no capability of 
compressed sensing. On the other hand, our RS analysis has clarified 
that the $L_1$-based scheme does have a considerably high 
reconstruction ability.
%
Moreover, it has been recognized that the $L_1$-reconstruction
can be solved via linear programming with a feasible computational
cost. These properties are advantageous 
from the viewpoint of practical utility. 

In this Letter, we have assumed that each entry of the
compression matrix $F$ is 
an i.i.d.\ random variable with 
zero mean and a fixed variance. 
Utilizing a technique offered in 
\cite{TakedaUdaKabashima}, the analysis can be extended to 
cases in which $F$ is randomly generated so as to be 
characterized as
\begin{eqnarray}
F^{\rm T} F = O D O^{\rm T}, 
\label{randomorthogonal}
\end{eqnarray}
where $D$ is a diagonal matrix, whose eigenvalue spectrum 
asymptotically converges to a fixed distribution,
and $O$ is a sample from the uniform distribution of 
$N \times N$ orthogonal matrices, independent of $D$. 
However, 
as long as $P \times P$ matrix $FF^{\rm T}$ is typically of full rank, 
which is the case when entries
of $F$ are i.i.d.\ random numbers of zero mean 
and a fixed variance, 
the result is identical to that obtained 
here (see \ref{randommatrix}). 
One can also show that 
the values of $\alpha_c(\rho)$ do not depend on 
details of the distribution of the non-zero elements of $\bx$
as long as the mean and variance are finite. 
This implies that the findings of this Letter generally hold for 
relatively wide classes of compression matrices
and signals.

Performance assessment of the $L_0$-reconstruction based 
on a replica symmetry breaking ansatz and 
development of mean field algorithms
for approximately solving the reconstruction problems 
with a lower computational cost are currently under way. 

{\em Note added --}
After submitting this Letter, the authors noticed that 
the typical criticality of the $L_1$-reconstruction 
was explored in \cite{Donoho2006,DonohoTanner2009}
for compression matrices consisting of i.i.d.\ 
zero mean Gaussian random column vectors 
utilizing techniques of {\em combinatorial geometry}.
It turns out that their {\em weak threshold} 
corresponds to our result for $\alpha_c(\rho)$ with $p=1$.  
In view of their criterion of reconstruction success, 
in which no errors are allowed, our result implies that 
the criticality of 
$L_1$-reconstruction is ``tight'' 
in the sense that it does not change irrespective 
of whether or not we allow small errors which are vanishing 
asymptotically as $N\to\infty$.  
The connection between our analysis and theirs further 
suggests a possibility of wide application of statistical-mechanics tools 
to problems in large-dimensional random combinatorial geometry.

\ack
This work was partially supported 
by Grants-in-Aid for Scientific Research on
the Priority Area ``Deepening and Expansion of Statistical Mechanical
Informatics'' by the Ministry of Education, Culture, Sports, Science
and Technology, Japan. 

\appendix

\section{Derivation of eq.~(\ref{freeenergy})}
\label{derivation}
Expressions
\begin{eqnarray}
\prod_{a=1}^n \delta(F\bbf^a-\bbs)
&=&\lim_{\tau \to +0} 
\frac{1}{\left (
\sqrt{2 \pi \tau}\right )^{nP}}
\exp \left [-\frac{1}{2 \tau}
\sum_{a=1}^n |F(\bbf^a-\bbf^0)|^2 \right ] \label{replica_delta1} \\
&=& 
\lim_{\tau \to +0} 
\frac{1}{\left (
\sqrt{2 \pi \tau}\right )^{nP}}
\exp \left [-\frac{1}{2 \tau}
\sum_{\mu=1}^P \sum_{a=1}^n (u_\mu^a-u_\mu^0)^2 \right ]
\label{replica_delta2}
\end{eqnarray}
play a key role in deriving eq.~(\ref{freeenergy}), 
where $\by=F\bx^0$ is used and 
$u_\mu^a = \sum_{i=1}^N F_{\mu i} x_i^a$ $(\mu=1,2,\ldots,P;\;
a=0,1,2,\ldots,n)$. 
When $F_{\mu i}$ are i.i.d.\ Gaussian random variables of 
mean zero and variance $1/N$, which is mainly assumed in this Letter, 
the central limit theorem guarantees that 
$u_\mu^a$ can be handled as zero mean 
multivariate Gaussian random variables which are characterized 
by the covariances 
$\left [u_\mu^a u_\nu^b \right ]_F=Q_{ab}\delta_{\mu \nu} $
for a fixed set of $\bbf^0,\bbf^1,\bbf^2,\ldots,\bbf^n$, 
where $\left [\cdots \right ]_F$ denotes the operation of 
averaging with respect to $F$, and $Q_{ab}=Q_{ba}=
N^{-1}\bbf^a \cdot \bbf^b$. 
$\delta_{\mu \nu}$ is unity for $\mu=\nu$ and vanishes otherwise. 
Under the RS ansatz 
\begin{eqnarray}
Q_{ab}=\left \{
\begin{array}{ll}
Q, & (a=b=1,2,\ldots,n) \cr
q, & (a>b=1,2,\ldots,n; \ b>a=1,2,\ldots,n) \cr
m, & (a=1,2,\ldots,n,b=0; \ a=0, b=1,2,\ldots,n) \cr
\rho, & (a=b=0)
\end{array}
\right . 
\label{RSorderparam}
\end{eqnarray}
this indicates that $u_\mu^a$ can be expressed as 
$u_\mu^a=\sqrt{Q-q} s_\mu^a + \sqrt{q} t_\mu$  
$(a=1,2,\ldots,n)$ and 
$u_\mu^0=\sqrt{\rho-m^2/q} s_\mu^0+m/\sqrt{q} t_\mu$, where
$s_\mu^a$ and $t_\mu$ 
are i.i.d.\ Gaussian random variables of zero mean and unit variance. 
Employing these expressions to eq.~(\ref{replica_delta2}) yields
\begin{eqnarray}
&&\frac{1}{N}\ln \left [\prod_{a=1}^n \delta(F\bbf^a-\bbs)\right ]_F
=\lim_{\tau \to +0}
\frac{1}{N}\ln 
\left (
\frac{\int Dv \exp \left [
\frac{n(q-2m+\rho)}{2(\tau+Q-q)}v^2 \right]}
{(2\pi)^{n/2}(\tau+Q-q)^{n/2}} \right )^{P} \cr
&&= -\frac{\alpha}{2}
\ln \left (\left (1-\frac{n(q-2m+\rho)}{Q-q} \right )(Q-q)^n 
(2\pi)^n \right )
\equiv {\cal T}_n(Q,q,m). 
\label{delta_average}
\end{eqnarray}
On the other hand, the saddle point method
offers an expression for the volume of the subshell 
corresponding to the RS order parameters (\ref{RSorderparam}) as
\begin{eqnarray}
&&\frac{1}{N} \left [
\int \prod_{a=1}^n
d \bbf^a e^{-\beta ||\bbf^a||_p}
{\cal I}(\{\bbf^a\}_{a=1}^n,\bx^0;Q,q,m)
\right ]_{\bbf^0} \cr
&&=\mathop{\rm extr}_{\widetilde{Q},\widetilde{q},\widetilde{m}}
\left \{
\frac{n \widetilde{Q}Q}{2}-\frac{n(n-1)\widetilde{q}q}{2}-n 
\widetilde{m}m
+\ln \Xi_n(\widetilde{Q},\widetilde{q},\widetilde{m};\beta)
\right \}\cr
&&\equiv {\cal S}_n(Q,q,m), 
\label{replica_entropy}
\end{eqnarray}
where
${\cal I}(\{\bbf^a\}_{a=1}^n,\bx^0;Q,q,m)=
\prod_{a=1}^n \left (\delta(|\bx^a|^2-NQ)
 \delta(\bx^0\cdot\bx^a-Nm) \right )
\prod_{a>b} \delta(\bx^a \cdot \bx^b-Nq)$ and
$
\Xi_n(\widetilde{Q},\widetilde{q},\widetilde{m};\beta)
=\lim_{\epsilon \to +0} \int Dz
\left [
\left (\int dx 
\exp \left [
-(\widetilde{Q}+\widetilde{q})x^2/2+(\sqrt{\widetilde{q}}z
+\widetilde{m}x^0)x
\right . \right . \right . $
$\left . \left . \left . 
-\beta|x|^{p+\epsilon} 
\right ] \right )^n \right ]_{x^0}
$. $\left [ \cdots \right ]_{x^0}=(1-\rho)\int dx^0 \delta(x^0)
(\cdots)+\rho \int Dx^0 (\cdots)$. 
Eqs.~(\ref{delta_average}) and (\ref{replica_entropy}) indicate
that $N^{-1}\ln \left [ Z^n(\beta;\by) \right ]=\mathop{\rm extr}_{Q,q,m}
\left \{ {\cal T}_n(Q,q,m)+{\cal S}_n(Q,q,m) \right \}$ 
holds for $n=1,2,\ldots$. 
For $\beta \to \infty$, the relevant variables scale so as to keep 
$\beta(Q-q)\equiv\chi$, 
$\beta^{-1} (\widetilde{Q}+\widetilde{q}) \equiv \widehat{Q}$, 
$\beta^{-2}\widetilde{q} \equiv \widehat{\chi}$
and $\beta^{-1}\widetilde{m} \equiv \widehat{m}$ of the order of unity. 
Analytically continuing the expression of 
$N^{-1}\ln \left [ Z^n(\beta;\by) \right ]$ from $n =1,2,\ldots $ to 
$n \in \mR$, in conjunction with employment of new variables
$\chi, \widehat{Q}, \widehat{\chi}$ and $\widehat{m}$, 
yields eq.~(\ref{freeenergy}). 

\section{Treatment of rotationally-invariant matrix ensembles}
\label{randommatrix}
For $\alpha =P/N \le 1$, let us suppose that compression matrix $F$ 
is characterized as eq.~(\ref{randomorthogonal}), where 
eigenvalues of diagonal matrix $D$ asymptotically 
follow a fixed distribution $r(\lambda)=(1-\alpha)\delta(\lambda)
+\alpha \widetilde{r} (\lambda)$ as $N,P \to \infty$ with keeping 
$\alpha=P/N \sim O(1)$ and $O$ is sampled from the uniform
distribution of $N \times N$ orthogonal matrices. 
This matrix ensemble is invariant under any rotation of coordinates. 
For simplicity, we assume that the support of $\widetilde{r}(\lambda)$ is 
a certain finite range away from the origin $\lambda = 0$, which implies 
that $P$ row vectors of a typical sample of $F$ are 
linearly independent. 
Under the RS ansatz, employment of 
a formula developed in \cite{TakedaUdaKabashima} to 
eq.~(\ref{replica_delta1}) offers an expression
\begin{eqnarray}
\frac{1}{N}\ln \left [\prod_{a=1}^n \delta(F\bbf^a-\bbs)\right ]_F
&=&\lim_{\tau \to +0}\left \{
G\left (-\frac{Q-q-n(Q-2m+\rho)}{\tau} \right ) \right .  \cr
&& 
\left . +(n-1)G\left (-\frac{Q-q}{\tau} \right )
-\frac{n\alpha}{2} \ln (2 \pi \tau) \right \}, 
\label{general_replica}
\end{eqnarray}
where
\begin{eqnarray}
G(-x)=\mathop{\rm extr}_{\Lambda}
\left \{
-\frac{1}{2}\int d\lambda {r} (\lambda)
\ln (\Lambda+\lambda)+\frac{\Lambda x}{2} \right \}
-\frac{1}{2}\ln x-\frac{1}{2}, 
\label{G-func}
\end{eqnarray}
for $x>0$. 
For $x \gg 1$, extremization in eq.~(\ref{G-func}) provides
$\Lambda \simeq (1-\alpha)/x$, which leads to an asymptotic form 
$G(-x) \simeq -(\alpha/2)\ln x-(\alpha/2)(1+\int d\lambda \widetilde{r} 
(\lambda) \ln \lambda)-((1-\alpha)/2)\ln (1-\alpha)$. 
Utilizing this expression in 
assessment of eq.~(\ref{general_replica}), 
where contributions of $O(\ln \tau)$ which arise from the $G$-functions 
are canceled with the term of $-(n\alpha/2)\ln \tau$
avoiding divergence as $\tau \to +0$, 
indicates that the difference between 
eqs.~(\ref{general_replica}) and (\ref{delta_average})
is only a constant 
independently of $\beta$. 
This means that in the vanishing temperature limit as $\beta \to \infty$
free energy for the rotationally-invariant ensemble 
exactly accords with eq.~(\ref{freeenergy}). 
Therefore, the result is identical to that obtained
in the main part of this Letter.

\end{document}